\documentclass[manuscript,nonacm]{acmart}
\AtBeginDocument{%
  }

\setcopyright{acmlicensed}
\copyrightyear{2026}
\acmYear{2026}
\acmDOI{XXXXXXX.XXXXXXX}
\acmConference[TOSEM 2026]{ACM Transactions on Software Engineering and Methodology}{August 1--2,
  2026}{place, HOLDER}
\acmISBN{978-1-4503-XXXX-X/2018/06}

\usepackage{caption}
\usepackage{subcaption}
\usepackage{amsmath}
\usepackage{multirow}
\usepackage{xcolor}
\usepackage[most]{tcolorbox}
\usepackage{algorithm}
\usepackage{algpseudocode}
\usepackage{hyperref}
\usepackage{graphicx}
\begin{document}

\title{FASE: Fast Adaptive Semantic Entropy for Code Quality}

\author{Shizhe Lin}
\affiliation{%
  \institution{University of Waterloo}
  \city{Waterloo}
  \country{Canada}}
\email{s222lin@uwaterloo.ca}
\orcid{0009-0005-2766-0281}

\author{Ladan Tahvildari}
\affiliation{%
  \institution{University of Waterloo}
  \city{Waterloo}
  \country{Canada}}
\email{ltahvild@uwaterloo.ca}
\orcid{0000-0001-8314-5560}

\renewcommand{\shortauthors}{Lin et al.}

\begin{abstract}

Multi-agent code generation offers a promising paradigm for autonomous software development by simulating the human software engineering lifecycle. However, system reliability remains hindered by LLM hallucinations and error propagation across interacting agents. While semantic entropy provides a principled way to quantify uncertainty without ground-truth answers, current methods often rely on costly LLM-driven equivalence checks. In this work, we introduce Fast Adaptive Semantic Entropy (FASE), a novel metric that approximates functional correctness based on the minimum spanning tree of structural and semantic dissimilarity graphs. Evaluations on HumanEval and BigCodeBench demonstrate that FASE outperforms state-of-the-art semantic entropy by LLM entailment, achieving a 25\% average improvement in Spearman correlation and a 19\% increase in ROCAUC score against Pass@1 from ground-truth test cases when using the Qwen3-Embedding-8B model. Furthermore, by eliminating costly LLM-driven equivalence evaluation, FASE incurs negligible computational overhead, requiring only approximately 0.3\% of the runtime cost of traditional semantic entropy approaches. These results position FASE as a practical, cost-effective solution for optimizing uncertainty quantification in real-world multi-agent workflows.

\end{abstract}

\begin{CCSXML}
<ccs2012>
   <concept>
       <concept_id>10002950.10003712</concept_id>
       <concept_desc>Mathematics of computing~Information theory</concept_desc>
       <concept_significance>500</concept_significance>
       </concept>
   <concept>
       <concept_id>10011007.10011074.10011092.10011782</concept_id>
       <concept_desc>Software and its engineering~Automatic programming</concept_desc>
       <concept_significance>500</concept_significance>
       </concept>
   <concept>
       <concept_id>10010147.10010178</concept_id>
       <concept_desc>Computing methodologies~Artificial intelligence</concept_desc>
       <concept_significance>500</concept_significance>
       </concept>
 </ccs2012>
\end{CCSXML}

\ccsdesc[500]{Mathematics of computing~Information theory}
\ccsdesc[500]{Software and its engineering~Automatic programming}
\ccsdesc[500]{Computing methodologies~Artificial intelligence}

\keywords{Generative AI, Agentic Code Generation, Adaptive Workflow, Trust in Agent Output}

\received{5 June 2026}

\maketitle

\section{Introduction}

Recent advances in large language models (LLMs) have accelerated the emergence of autonomous multi-agent systems for software engineering tasks~\cite{ferrag2025llm,liu2024large,he2025llm}. Rather than relying on a single model to generate code end-to-end, modern frameworks increasingly decompose software development into collaborative workflows involving specialized agents responsible for requirement analysis, planning, coding, testing, debugging, and review~\cite{talebirad2023multi,dong2024self}. Inspired by human software engineering practices, systems such as MetaGPT~\cite{hong2024metagpt}, CodeCoR~\cite{pan2025codecor}, and AdaCoder~\cite{zhu2025adacoder} demonstrate that role specialization and structured collaboration can substantially improve code generation quality and task-solving capability for complex programming problems. These agentic workflows further enable LLMs to address repository-level development tasks, long-horizon reasoning, and iterative refinement processes that are difficult for standalone models to handle effectively.

\begin{figure}[h]
    \centering
    \includegraphics[width=0.8\linewidth]{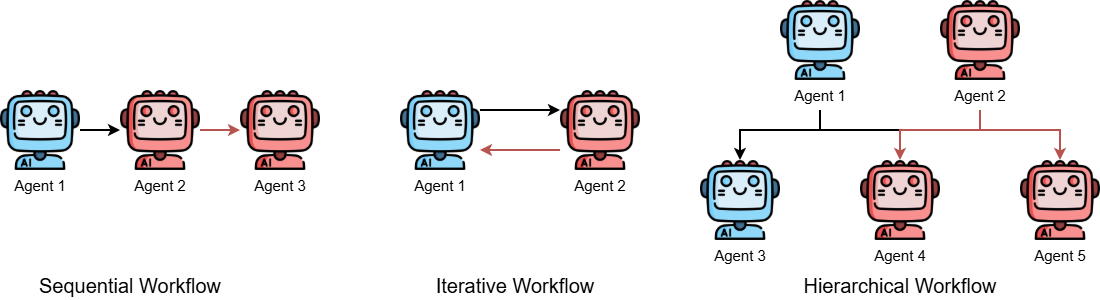}
    \caption{The risk of hallucination from unreliable agents (red) undermines the credibility of coherent agents (blue) and propagates across different workflow architectures.}
    \label{fig:error_propagation}
\end{figure}

Despite these advances, LLM-based software engineering systems remain fundamentally limited by hallucinations and uncertainty in generated outputs~\cite{liu2024exploring,agarwal2024codemirage,liu2026beyond,yehudai2025survey}. Errors produced during early stages of reasoning or implementation can propagate across agents, leading to cascading failures throughout the development pipeline~\cite{mohammadi2025evaluation} as shown in Fig.~\ref{fig:error_propagation}. Existing approaches attempt to mitigate these issues through self-reflection~\cite{kadavath2022language}, retrieval-augmented generation~\cite{eghbali2024hallucinator,das2025multi}, iterative debugging~\cite{ji2023towards,fakhoury2024llm}, static analysis~\cite{yang2025advancing,liu2024refining}, mutation testing~\cite{wang2026mutation,tip2025llmorpheus}, and fuzz testing~\cite{chen2025traceawareness,cao2025program}. Recent frameworks also introduce consensus-based reasoning~\cite{xu2026hallucination}, and repository-aware grounding strategies~\cite{liao2024mathbf} to improve factual consistency and functional correctness. 

However, many of these techniques rely on expensive iterative interactions with LLMs, resulting in substantial computational overhead and limited scalability for real-world deployment. Entropy-based uncertainty estimation methods have recently been explored to evaluate the reliability of LLM-generated code without requiring ground-truth labels~\cite{kuhn2023semantic}. Semantic entropy~\cite{farquhar2024detecting} measures uncertainty by grouping outputs according to functional equivalence rather than textual similarity, but it requires bidirectional entailment checks using LLMs, which limits scalability in practical multi-agent systems. To improve efficiency, structural entropy~\cite{song2025measuring} has been proposed as an alternative designed specifically for source code. Nevertheless, structural similarity alone cannot fully capture program semantics, as functionally equivalent solutions may have different structures while structurally similar code may still exhibit different behaviours. Building upon the prior works, this paper makes the following contributions:

\begin{itemize}
    \item \textbf{Lightweight semantic entropy via embedding models:} Fast Adaptive Semantic Entropy (FASE) that leverages the semantic signals captured by code embedding models and dynamically adapts clustering according to the unique structure of each programming task's solution space. FASE enables the estimation of the functional correctness of LLM-generated code without requiring ground-truth test cases while incurring only negligible computational overhead.
    \item \textbf{Critical evaluation on estimation of code functional correctness:} Evaluation on the reliability and accuracy of the proposed FASE entropy against existing LLM-based semantic entropy methods, self-evaluation techniques, and other baseline approaches in terms of their correlation and predictive capability for code functional correctness.
    \item \textbf{Cost–accuracy trade-off analysis for multi-agent observation:} Quantitative comparison of computational runtime between FASE and other baselines, demonstrating that the proposed approaches offer substantially improved efficiency while maintaining useful correctness signals for scalable multi-agent systems.
\end{itemize}

The rest of the paper is organized as follows. Section~\ref{sec:problem_statement} formalizes the problem and identifies the key research gaps. Section~\ref{sec:methodology} presents the methodology of the proposed FASE approach. Section~\ref{sec:evaluation} outlines the research questions addressed in this work, details the experimental setup, reports the results, and provides a comprehensive analysis of the evaluation. The related work and state-of-the-art are listed in Section~\ref{sec:related_work}. Section~\ref{sec:threat_to_validity} discusses threats to validity. Finally, Section~\ref{sec:conclusion} concludes by summarizing the main findings and discussing future research directions.

\section{Problem Definition}
\label{sec:problem_statement}
The increasing adoption of LLM systems for software development creates a growing need for efficient and reliable methods to estimate the quality of generated code without relying on ground-truth test cases.

\noindent{\textbf{Entropy of LLM Output:}} One common measure of uncertainty is the predictive entropy of the output distribution, which quantifies the amount of information the model has about the output given the input~\cite{lindley1956measure}: \(PE(X)=H(Y|x)=-\sum P(y|x) \ln P(y|x)\) where the predictive entropy for an input \(x\) is defined as the conditional entropy of the output random variable \(Y\) given \(x\). Low predictive entropy reflects a sharply peaked output distribution, implying high model confidence, while high predictive entropy indicates a more uniform distribution in which many possible outputs have comparable likelihoods.

The derivation of entropy for a set of code samples \(X\) given a programming task is presented in Eq.~\ref{eq:class_entropy}, where \(c\) denotes an equivalence class within the set of all possible classes \(\mathcal{C}\). In practice, however, only a finite set of code samples is available, and the computation should be based on individual output \(s\) observed within the sampled set.

\begin{equation}
PE(X)=H(\mathcal{C}|x)=-\sum_c P(c|x)\log P(c|x)=-\sum_c\left( \left[ \sum_{s\in c}P(s|x)\right]\log \left[ \sum_{s\in c}P(s|x)\right]\right)
\label{eq:class_entropy}
\end{equation}

\noindent{\textbf{Textual, Structural and Semantic Equivalence:}} The equivalence classes \(c\) of outputs are determined by clustering the results using an equivalence function \(E(\cdot,\cdot)\). For a given equivalence class, all output sentences within the class are pairwise equivalent, \(\forall s,s'\in c:E(s,s')\), meaning they convey the same content. 

A straightforward approach to determining whether two outputs are equivalent is to compare their textual similarity but it fails to reliably distinguish between coherent and hallucinated responses~\cite{farquhar2024detecting,song2025measuring}. Structural entropy~\cite{song2025measuring} addresses this limitation by introducing a higher-level perspective that abstracts away trivial textual differences and groups code based on similarities in their abstract syntax trees. Nevertheless, solutions to programming tasks are not constrained to a single structural form, and functionally equivalent programs may exhibit diverse syntactic representations. To overcome this, recent work has proposed semantic entropy~\cite{farquhar2024detecting} through LLM-based bidirectional entailment, which quantifies uncertainty at the level of meaning rather than tokens, providing a more faithful estimate of a model’s confidence. However, applying this approach directly to programming tasks remains imperfect~\cite{maveli2024can}, as LLMs often fail to detect subtle code variations that can result in substantial differences in functionality.

There are three limitations of the current state-of-the-art entropy calculation:
\begin{itemize}
    
    \item The accuracy of bidirectional entailment relies on LLM's self-evaluation capability; different models often produce inconsistent judgments, making it difficult to identify a reliable and universally optimal evaluator, particularly in the coding domain.
    
    \item The cost of producing equivalence-class labels grows exponentially with the number of generated outputs for LLM-based semantic entropy, rendering the method impractical for multi-step, multi-agent workflows where time efficiency is critical.
    
    \item Structural entropy alone fails to account for multiple valid structural representations of a given task, often assigning functionally equivalent code samples to different classes.
\end{itemize}

These limitations motivate the need for more accurate, efficient, and model-agnostic approaches to estimate the consistency of functionality in code generation.

\section{Methodology}
\label{sec:methodology}

\begin{figure}[h]
    \centering
    \includegraphics[width=0.8\linewidth]{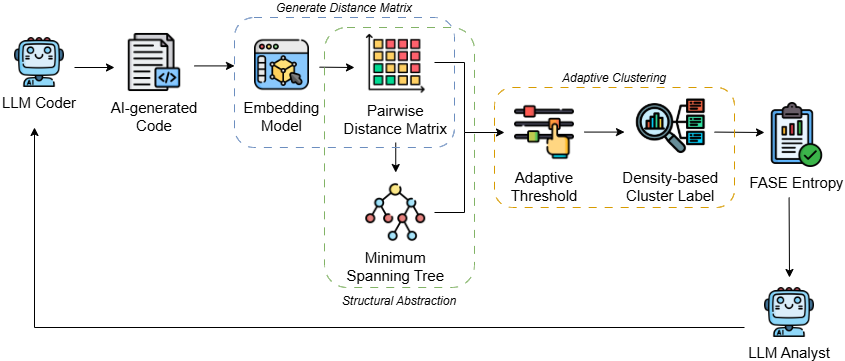}
    \caption{The workflow of computing encoder-based semantic entropy for the samples of code generated for a given task.}
    \label{fig:workflow}
\end{figure}

The proposed FASE entropy for LLM-generated code offers a more efficient approach to computing semantic entropy and estimating code quality without relying on equivalence prediction of LLMs, following the workflow shown in Fig.~\ref{fig:workflow}\footnote{Fig.1 and Fig.2 contain icons designed using resources from Flaticon.com}. By mapping code artifacts into a continuous embedding space, their semantic relationships can be quantified directly, enabling fast and cost-effective assessment of functional similarity and divergence among generated solutions.

\noindent{\textbf{Pairwise Semantic Distance via Encoder-Only Embedding Models:}} The current realization of the semantic equivalence function \(E(\cdot,\cdot)\) relies on LLM-based bidirectional entailment~\cite{farquhar2024detecting}. While effective, this approach incurs substantial computational cost from LLM inference. This combination of high per-query cost and \(\mathcal{O}(|x|^2)\) scaling results in significant overhead, rendering the method impractical for deployment in real-world or large-scale production settings. This paper leverages the low computational cost and high efficiency of embedding models to replace the existing approach, enabling finer-grained estimation of the pairwise semantic distance matrix at substantially reduced cost. For the samples of AI-generated code \(x\) from the LLM coder, their pairwise semantic distance matrix consists of \(|x|^2\) values where \(d_{i,j} \in[0,1]\) represents the semantic difference between the generated code \(x_i\) and \(x_j\), \(i,j \in |x|\) through cosine distance. As \(|x|\) increases, the estimated distribution converges to the true underlying distribution. The resulting class labels can then be used to compute semantic entropy in the same manner as approaches that rely on LLMs to judge bidirectional functional equivalence.

\noindent{\textbf{Structural Abstraction using Minimum Spanning Trees:}} While the full matrix encodes all pairwise distances, it is often redundant and obscures local structure due to its density. Extracting a minimum spanning tree (MST) from a densely connected pairwise distance matrix provides a compact representation of the underlying semantic structure by preserving only the most informative relationships among samples~\cite{gagolewski2025clustering}. The MST retains the subset of edges \((x_i,x_j,w)\) for \(x_i,x_j\) with \(i,j \in |x|\) of weight \(w\) that connect all samples with minimal total distance, effectively highlighting the strongest semantic affinities while discarding weaker, less informative connections. As a result, the MST reveals the intrinsic topology of the data, including local neighborhoods, cluster boundaries, and the relative dispersion of samples. This abstraction enables efficient downstream analysis while preserving essential structural information.

\noindent{\textbf{Adaptive Density-Based Clustering for Equivalence Class Construction:}} With access to a pairwise distance matrix, density-based methods can conveniently be used to compute equivalence clusters. However, fixed reachability or cluster size do not generalize well across tasks since the number and size of semantic equivalence classes are precisely the quantities we aim to estimate for entropy computation. This issue can be addressed by our adaptive density-based clustering algorithm, which leverages key structural properties derived from the pairwise distance matrix and the extracted minimum spanning tree (MST). Empirically, we observe that: (1) even when two sample sets exhibit similar average pairwise distances, those containing meaningful local clusters will yield an MST with a smaller average edge weight; and (2) edges that connect distinct clusters tend to have significantly larger weights than edges within the same cluster. Based on these observations, the adaptive neighborhood threshold \(\epsilon\) is determined using the \textit{mode} of the MST edge weights \(w\), thereby reducing the influence of the relatively rare, high-weight inter-cluster edges. The mode is estimated as the maximizer of a Gaussian kernel density estimation computed over \(w\). The \(\mathrm{Mo}(w)\) is then scaled by the ratio between the mean edge weight of the MST and the mean value of the full distance matrix as shown in Eq.~\ref{eq:adaptive_epsilon}:
\begin{equation}
\epsilon_{adaptive}=\alpha\cdot\mathrm{Mo}(w)\cdot(1-\frac{\bar{w}}{\bar{d}})
\label{eq:adaptive_epsilon}
\end{equation}
where \(\alpha\) is a constant used to control the clustering granularity. The choice of \(\alpha\) depends on the selected coder LLM and embedding model across the entire code generation dataset, unlike \(\bar{w}\) or \(\bar{d}\), which are task-specific adaptive variables. The semantic equivalence class labels of the generated codes become FASE entropy through Eq.~\ref{eq:class_entropy}.

To summarize, FASE introduces a novel approach for uncertainty estimation in code generation by replacing expensive LLM-based semantic equivalence checks with an efficient embedding-driven graph analysis framework. By combining semantic embeddings, minimum spanning tree extraction, and adaptive density-based clustering, FASE captures functional uncertainty without ground-truth knowledge and further bridges the gap between fast but structurally limited syntax-based approaches and semantically accurate but computationally expensive LLM-based evaluation, providing a practical and effective solution for reliable code quality estimation for large multi-agent workflows and real-world software engineering scenarios.

\section{Evaluation}
\label{sec:evaluation}

The evaluation section begins by presenting the research questions, followed by a description of the experimental setup. It then investigates the properties of the PDM and MST in differentiating functional variations among generated code samples. Next, the study demonstrates the necessity of adaptive clustering for semantic entropy estimation. Subsequently, the proposed FASE entropy and its variants are compared against existing baselines. Finally, the section provides a comprehensive analysis of the overhead associated with all evaluated metrics.

The development of the proposed FASE entropy aims to answer the following research questions:
\begin{itemize}

    \item \textbf{RQ1.} \textit{How effectively can embedding-based graph abstractions in the form of Pairwise Distance Matrix (PDM) and Minimum Spanning Tree (MST) differentiate between tasks with high and low functional consistency in code generation?} This research identifies the most influential factors derived from the pairwise distance matrix and the extracted MST that effectively differentiate the quality of generated code samples.

    \item \textbf{RQ2.} \textit{To what extent does task-specific adaptive clustering optimize the modeling of semantic equivalence classes compared to fixed-threshold baselines?} Given the distance matrix, density-based clustering methods can efficiently group code samples into equivalence classes; however, their performance is highly sensitive to key hyperparameters. Whether a fixed criterion can generalize across different coding tasks remains an open question that requires empirical evaluation.

    \item \textbf{RQ3.} \textit{How do FASE and its structural hybrid variants compare against state-of-the-art LLM-based semantic entropy and self-evaluation baselines in predicting code functional correctness within multi-agent workflows?} To demonstrate the effectiveness of the proposed method, experiments are conducted comparing it against state-of-the-art structural and semantic entropy approaches, as well as other self-evaluation methods.
    
    \item \textbf{RQ4.} \textit{What are the computational efficiency and scalability gains of FASE compared to traditional LLM-driven bidirectional entailment checks?} A quantitative comparison of runtime and resource usage is required to demonstrate the practical value of alternative observation methods.
    
\end{itemize}

Answering these research questions provides a comprehensive understanding of both the effectiveness and efficiency of embedding-based uncertainty estimation for code generation. Together, these analyses establish a principled trade-off between accuracy and efficiency, demonstrating the practicality of FASE for scalable code quality assessment in real-world multi-agent systems.

\subsection{Experiment Setup}

Two code generation benchmarks are selected to examine the correlation between semantic entropy and code generation quality. 
\begin{itemize}
  \item HumanEval~\cite{zheng2023codegeex} is widely used as a foundational benchmark for code-generation models. It consists of 164 handwritten Python programming problems of basic algorithms. 
  \item BigCodeBench~\cite{zhuo2024bigcodebench} is a more challenging and practical code-generation benchmark compared to HumanEval. In particular, we use the BigCodeBench-hard subset, which contains 148 Python problems that are closer to real-world programming scenarios.
\end{itemize}

Four open-source 7-billion-parameter LLMs are selected as coders and analysts in the experiments: Mistral-7B~\cite{jiang2023mistral}, CodeLlama-7B~\cite{roziere2023code}, DeepSeek-Coder-7B~\cite{guo2024deepseek}, and Qwen2.5-Coder-7B~\cite{hui2024qwen2}. By evaluating this diverse set of models, this research explores the generality of semantic entropy signals across different model designs, training objectives, and data distributions. The number of parameters is identical to avoid the impact of model scale.

\begin{table}[htbp]
\centering
\caption{Embedding Models Used in the Study}
\label{tab:embedding_models}

\begin{tabular}{|l|ccc|}
    \hline
    \textbf{Name} & \textbf{\# Parameters} & \textbf{Dimensionality} & \textbf{Context Window} \\
    \hline
    All-MiniLM-L6-v2 & 0.02B & 384 & 0.2K \\
    GTE-ModernBERT-base & 0.1B & 768 & 8K \\
    Llama-Embed-Nemotron & 8B & 4096 & 32K \\
    \hline
    Qwen3-Embedding & 0.6B & 1024 & 32K \\
    Qwen3-Embedding & 4B & 2560 & 32K \\
    Qwen3-Embedding & 8B & 4096 & 32K \\
    \hline
\end{tabular}

\end{table}
For generating semantic embeddings of code and task descriptions, four encoder-only embedding models of different sizes are selected with high performance in text embedding, searching, ranking and clustering tasks as shown in Tab.~\ref{tab:embedding_models}: All-MiniLM-L6-v2~\cite{wang2020minilm}, GTE-ModernBERT-base~\cite{zhang2024mgte}, Llama-Embed-Nemotron~\cite{babakhin2025llamaembednemotron8buniversaltextembedding} and Qwen3-Embedding~\cite{qwen3embedding}. Their embedding dimensionalities are 384, 768 and 4096 for All-MiniLM, ModernBERT, and Nemotron, respectively. We also explored the impact of parameter size on the same model for Qwen3-Embedding of 0.6B, 4B and 8B parameters with dimensional vector size of 1024, 2560 and 4096. Each encoder differs in underlying training design and domain emphasis, allowing us to test whether semantic representation quality impacts semantic distance, clustering results, and ultimately the estimation of functional correctness. By incorporating this variety of embedding models, we aim to strengthen the soundness and generality of our experimental findings.

The experiment is conducted on ASUSTeK ESC4000A-E12 with AMD EPYC™ 9554 processor on a Ubuntu 24.04.2 LTS environment. The GPU we used is NVIDIA H100 NVL with driver version 550.163.01 and CUDA version 12.4. We deployed \textit{CodeLlama-7b-Instruct-hf} for the agent model. When generating output with nucleus sampling, we set the \textit{temperature} to 0.5 and \textit{top\_p} to 0.95 for more consistent results. For each task, 10 samples will be generated to compute related metrics. The implementation and experimental artifacts are publicly available at \href{https://github.com/corvolin/CSE4AgenticSoftDev}{this link}\footnote{https://github.com/corvolin/CSE4AgenticSoftDev}

\subsection{Effectiveness of Embedding Representations in Capturing Functional Correctness}

\begin{figure}[htbp]
    \centering
    \includegraphics[width=0.66\linewidth]{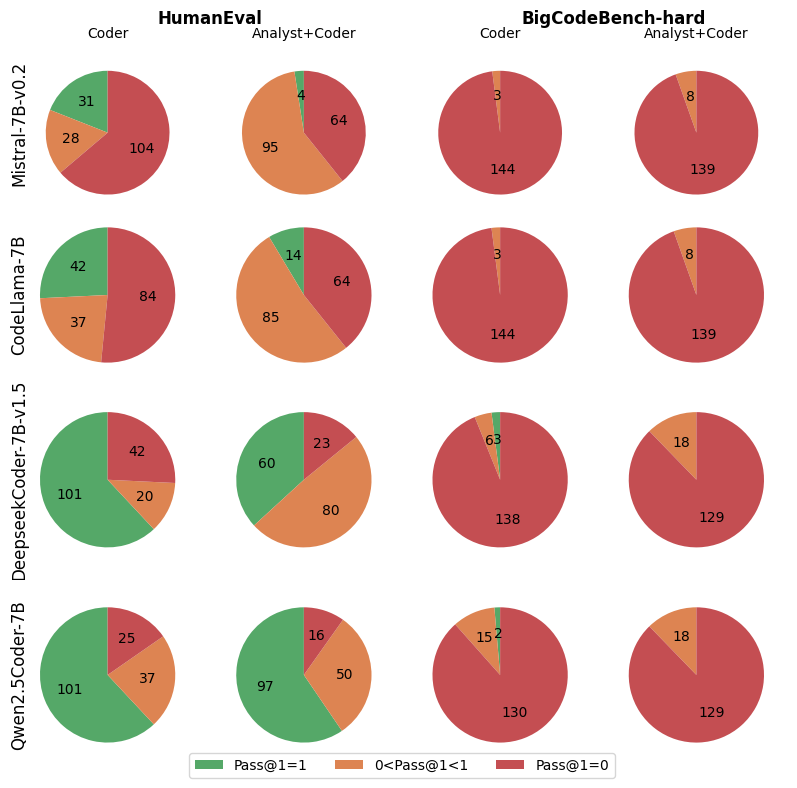}
    \caption{Pass@1 distribution of code generated by Mistral, CodeLlama, DeepseekCoder, Qwen2.5-Coder on HumanEval and BigCodeBench-hard}
    \label{fig:code_gen_result}
\end{figure}

The overall performance of the selected models on code generation benchmarks is summarized in Fig.~\ref{fig:code_gen_result}, which presents results on the HumanEval and BigCodeBench-hard tasks. 

The green region denotes the number of tasks with pass@1~\cite{chen2021evaluating} equal to 1, indicating that the model consistently generates functionally correct solutions on the first attempt. The red region represents tasks with pass@1 equal to 0, where no correct solutions are produced. The orange region captures intermediate cases with 0<pass@1<1, reflecting partial success across multiple attempts. 

On the HumanEval benchmark, most LLMs are able to generate partially correct solutions, with the number of tasks achieving pass@1 = 1 ranging from 31 for Mistral to 101 for both DeepSeek-Coder and Qwen2.5-Coder. In contrast, performance on the BigCodeBench-Hard benchmark is substantially lower: the majority of models fail to produce any functionally correct solutions. The best performance is achieved by Qwen2.5-Coder, with only 2 tasks reaching pass@1 = 1 and 15 tasks exhibiting partial success with 0<pass@1<1. We also evaluated the effect of introducing an analyst agent that provides semi-structured information on data and control flow based on the task’s functional requirements. The impact on pass@1 is mixed. Across both HumanEval and BigCodeBench-Hard, the number of tasks with pass@1 = 0 and pass@1 = 1 decreases, while the number of tasks with intermediate outcomes of 0<pass@1<1 increases. For instance, Qwen2.5-Coder reduces 9 cases of pass@1 = 0 at the cost of 4 cases of pass@1 = 1. In contrast, other models tend to exhibit a larger reduction in pass@1 = 1 than in pass@1 = 0. These results suggest that naively introducing an analyst agent without task-specific adaptation does not necessarily improve overall performance, aligning with growing concerns in the research community regarding hallucination propagation in multi-agent systems.

% font of x y axix
\begin{figure*}[htbp]
\centering

% Row 1
\begin{subfigure}[t]{0.23\textwidth}
    \centering
    \includegraphics[width=\linewidth]{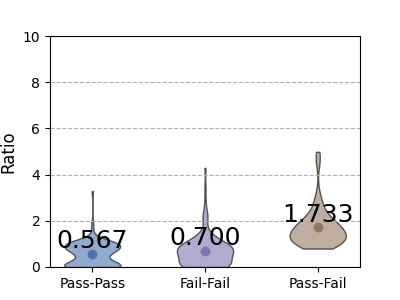}
    \caption{All-MiniLM PDM}
\end{subfigure}
\hfill
\begin{subfigure}[t]{0.23\textwidth}
    \centering
    \includegraphics[width=\linewidth]{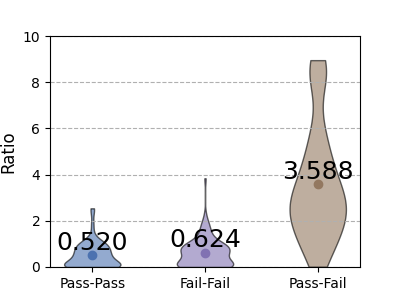}
    \caption{All-MiniLM MST}
\end{subfigure}
\hfill
\begin{subfigure}[t]{0.23\textwidth}
    \centering
    \includegraphics[width=\linewidth]{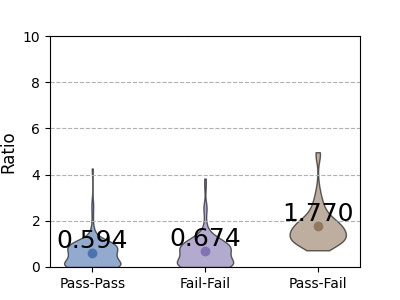}
    \caption{Qwen3-Emb.-0.6B PDM}
\end{subfigure}
\hfill
\begin{subfigure}[t]{0.23\textwidth}
    \centering
    \includegraphics[width=\linewidth]{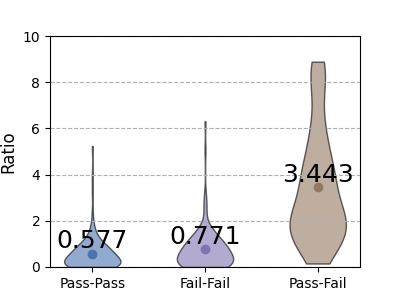}
    \caption{Qwen3-Emb.-0.6B MST}
\end{subfigure}

\vspace{0.5em}

% Row 2
\begin{subfigure}[t]{0.23\textwidth}
    \centering
    \includegraphics[width=\linewidth]{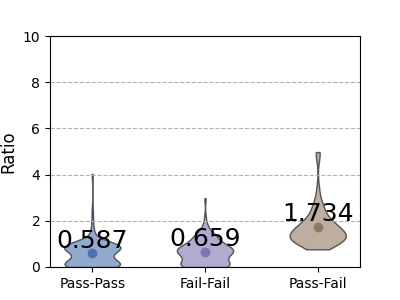}
    \caption{ModernBERT PDM}
\end{subfigure}
\hfill
\begin{subfigure}[t]{0.23\textwidth}
    \centering
    \includegraphics[width=\linewidth]{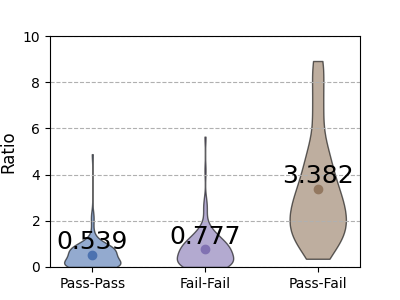}
    \caption{ModernBERT MST}
\end{subfigure}
\hfill
\begin{subfigure}[t]{0.23\textwidth}
    \centering
    \includegraphics[width=\linewidth]{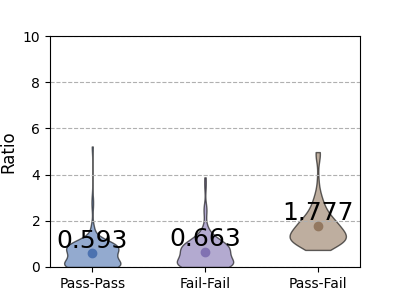}
    \caption{Qwen3-Emb.-4B PDM}
\end{subfigure}
\hfill
\begin{subfigure}[t]{0.23\textwidth}
    \centering
    \includegraphics[width=\linewidth]{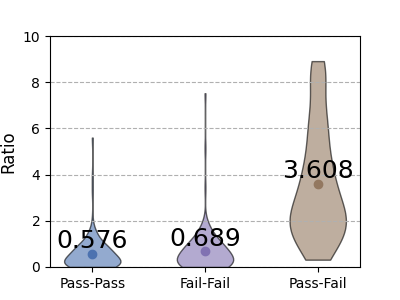}
    \caption{Qwen3-Emb.-4B MST}
\end{subfigure}

\vspace{0.5em}

% Row 3
\begin{subfigure}[t]{0.23\textwidth}
    \centering
    \includegraphics[width=\linewidth]{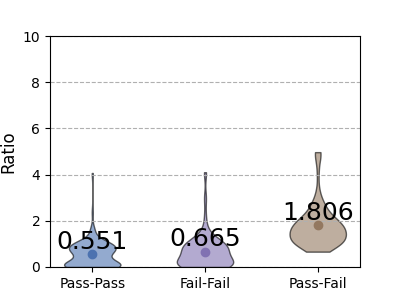}
    \caption{Nemoron PDM}
\end{subfigure}
\hfill
\begin{subfigure}[t]{0.23\textwidth}
    \centering
    \includegraphics[width=\linewidth]{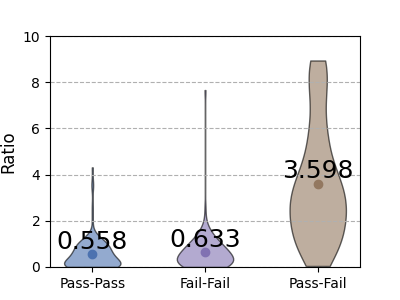}
    \caption{Nemotron MST}
\end{subfigure}
\hfill
\begin{subfigure}[t]{0.23\textwidth}
    \centering
    \includegraphics[width=\linewidth]{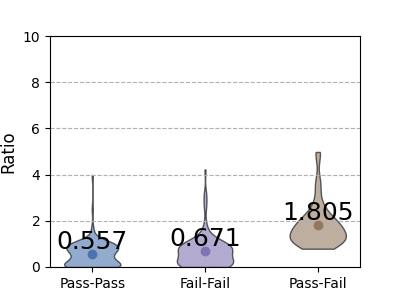}
    \caption{Qwen3-Emb.-8B PDM}
\end{subfigure}
\hfill
\begin{subfigure}[t]{0.23\textwidth}
    \centering
    \includegraphics[width=\linewidth]{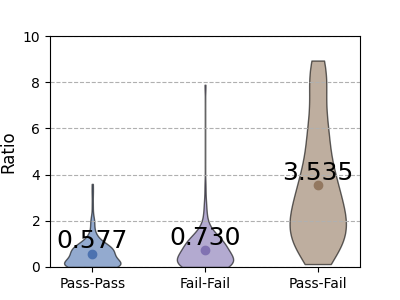}
    \caption{Qwen3-Emb.-8B MST}
\end{subfigure}

\caption{The ratio of mean distance in Pairwise Distance Matrix (PDM) and mean edge weight in MST categorized by their connected nodes, Pass-Pass(Blue), Fail-Fail(Purple) and Pass-Fail(Brown), compared to all distances and weights in its samples.}
\label{fig:edge_type_difference}

\end{figure*}

The PDM generated from the embedding models indeed capture meaningful semantic differences among code samples for each task. This research firstly focus on tasks with intermediate pass@1 values and categorize pairwise distances into three groups: (i) between functionally correct codes that passed provided test cases, (ii) between codes that failed the test cases, and (iii) between a correct and a failed code. For each category, the ratio between its mean distance and the overall mean distance of the PDM is computed, and the distribution of these ratios are shown in Fig.~\ref{fig:edge_type_difference}. Across all embedding models, distances between pairs of functionally correct codes are consistently the smallest. Distances between failed code pairs are slightly larger, likely reflecting diverse failed root cause despite shared task objectives. Notably, both categories yield ratios below 1, whereas distances between correct and failed code pairs consistently exceed 1. This indicates that substantial functional differences correspond to significantly larger semantic distances relative to the task-specific average.

After extracting the MST from the PDM, the previously observed pattern remains consistent. Edges connecting functionally correct codes exhibit the smallest ratio, followed by edges between failed codes. In contrast, edges linking correct and failed codes show substantially larger ratio, increasing from approximately 1.7–1.8 in the PDM to 3.4–3.6 in the MST. This amplification indicates that the MST preferentially selects low-weight edges that capture local structure, effectively forming tight clusters among semantically similar codes. Once these intra-cluster connections are established, the higher-weight edges representing connections between functionally distinct groups are incorporated to link the clusters. Consequently, the extracted MST further differentiates the separation of semantic clusters compared to the original PDM.

\begin{figure}[htbp]
\centering

% Row 1
\begin{subfigure}[t]{0.32\linewidth}
    \centering
    \includegraphics[width=\linewidth]{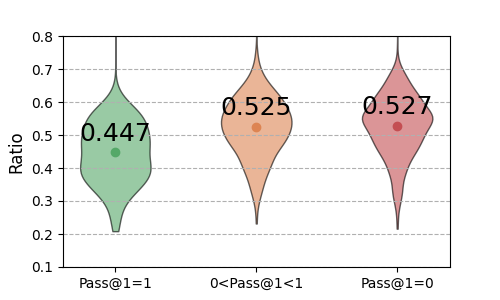}
    \caption{All-MiniLM}
\end{subfigure}
\hfill
\begin{subfigure}[t]{0.32\linewidth}
    \centering
    \includegraphics[width=\linewidth]{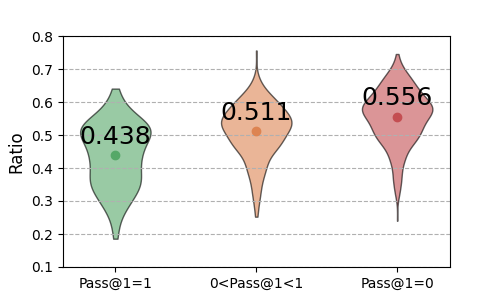}
    \caption{ModernBERT}
\end{subfigure}
\hfill
\begin{subfigure}[t]{0.32\linewidth}
    \centering
    \includegraphics[width=\linewidth]{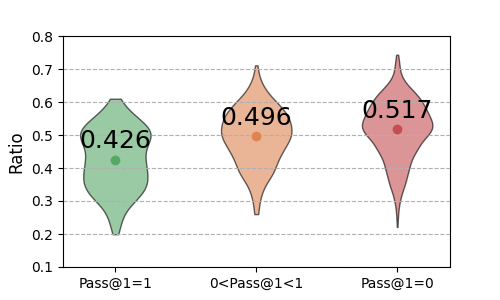}
    \caption{Nemotron}
\end{subfigure}

\vspace{0.5em}

% Row 2
\begin{subfigure}[t]{0.32\linewidth}
    \centering
    \includegraphics[width=\linewidth]{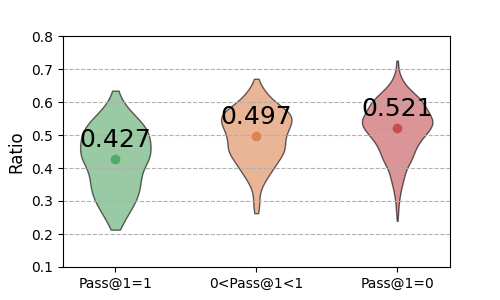}
    \caption{Qwen3-Emb.-0.6B}
\end{subfigure}
\hfill
\begin{subfigure}[t]{0.32\linewidth}
    \centering
    \includegraphics[width=\linewidth]{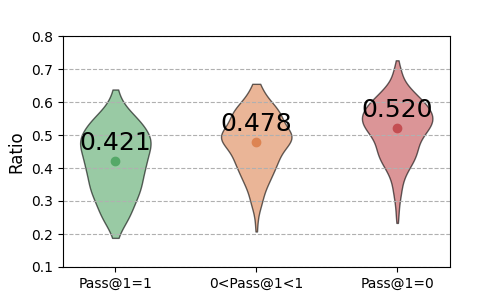}
    \caption{Qwen3-Emb.-4B}
\end{subfigure}
\hfill
\begin{subfigure}[t]{0.32\linewidth}
    \centering
    \includegraphics[width=\linewidth]{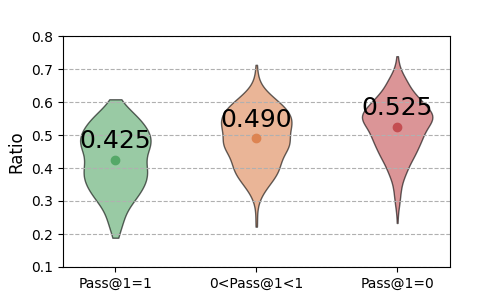}
    \caption{Qwen3-Emb.-8B}
\end{subfigure}

\caption{The ratio between mean distance in Pairwise Distance Matrix (PDM) and mean edge weight in MST categorized for tasks with Pass@1 of 0, between 0 and 1, and 1.}
\label{fig:mst_to_global_mean}
\end{figure}

However, as shown in Fig.~\ref{fig:code_gen_result}, the majority of code generation tasks across all four LLMs exhibit either pass@1 = 0 or pass@1 = 1. In such cases, there are few or no edges connecting functionally correct and failed codes with substantially higher weights relative to the task average. Moreover, the difference of intra-cluster edges between pairs of correct codes and pairs of failed codes is relatively small. Although not shown, the distributions of pairwise distances and MST edge weights do not differ significantly between tasks with pass@1 = 0 and those with pass@1 = 1.

While examining the PDM and MST in isolation does not reveal strong differences between tasks with pass@1 = 1 and pass@1 = 0, comparing edge weights before and after MST extraction provides additional insight. Fig.~\ref{fig:mst_to_global_mean} illustrates the ratio between the mean edge weight of the MST and the mean pairwise distance of the PDM for each task. Across all embedding models, a consistent pattern emerges: tasks with pass@1 = 1 exhibit the lowest ratios, followed by tasks with intermediate pass@1 values, while tasks with pass@1 = 0 have the highest ratios. This indicates that, although overall distance distributions appear similar across tasks, those with pass@1 = 1 contain stronger local structure, reflected by relatively smaller MST edge weights. In contrast, tasks with pass@1 = 0 lack such structure, with their minimum edges remaining comparable to the overall pairwise distances.

\begin{center} 
\begin{tcolorbox}[
  colback=gray!15,        % light gray background
  colframe=gray!50,       % gray frame
  arc=3mm,                % rounded corner radius
  width=0.9\textwidth,    % the box is 80% of text width
  left=4pt, right=4pt,    % horizontal padding
  top=4pt, bottom=4pt     % vertical padding
]

\textbf{Answer to RQ.1:} The cosine-distance distributions derived from the PDM and its corresponding MST capture meaningful patterns of functional consistency among generated code samples. In particular, high-weight edges in the extracted MST often correspond to genuine functional differences between correct and incorrect solutions. Although the overall distance distributions for tasks with Pass@1 = 1 and Pass@1 = 0 appear similar, tasks with Pass@1 = 1 consistently exhibit a substantially lower mean MST-to-PDM weight ratio.
\end{tcolorbox}
\end{center}

\subsection{Impact of Adaptive Clustering on Semantic Equivalence Modeling}

Given the pairwise distance matrix derived from semantic embeddings of the generated code, an intuitive approach for identifying semantic clusters is to apply density-based clustering techniques~\cite{ester1996density,campello2013density}. The foundation of adaptive clustering is DBSCAN with a fixed neighborhood radius \(\epsilon\) and HDBSCAN with a fixed minimum cluster size (MCS). The parameter \(\epsilon\) is varied from 0.01 to 0.2, covering the largest range of pairwise semantic distances across different embedding models. The MCS varies from 2 to 9, with a single cluster of size 10 permitted. The adaptive \(\epsilon\) based on the MST/PDM ratio are derived as Eq.~\ref{eq:adaptive_epsilon}. The constant \(\alpha\) ranges from 0.1 to 3 to control clustering granularity.

\begin{figure}[htbp]
\centering

% Row 1
\begin{subfigure}[t]{0.32\linewidth}
    \centering
    \includegraphics[width=\linewidth]{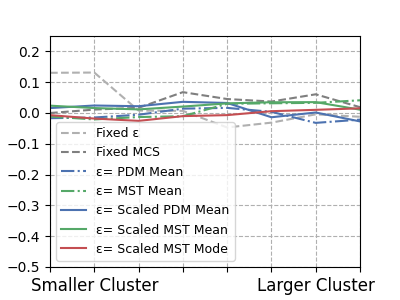}
    \caption{All-MiniLM}
\end{subfigure}
\hfill
\begin{subfigure}[t]{0.32\linewidth}
    \centering
    \includegraphics[width=\linewidth]{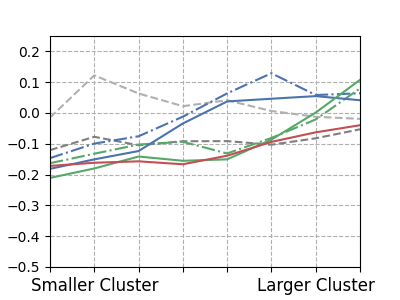}
    \caption{ModernBERT}
\end{subfigure}
\hfill
\begin{subfigure}[t]{0.3\linewidth}
    \centering
    \includegraphics[width=\linewidth]{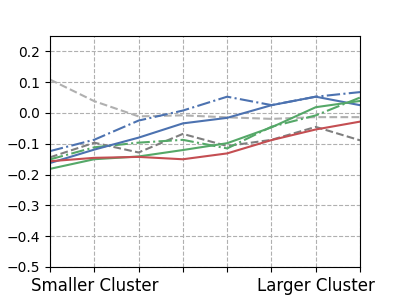}
    \caption{Nemotron}
\end{subfigure}

\vspace{0.5em}

% Row 2
\begin{subfigure}[t]{0.32\linewidth}
    \centering
    \includegraphics[width=\linewidth]{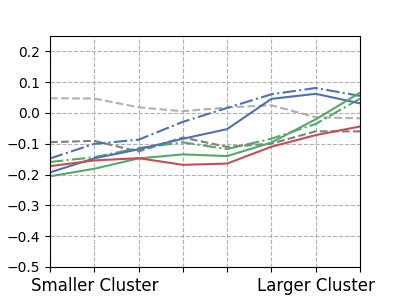}
    \caption{Qwen3-Emb.-0.6B}
\end{subfigure}
\hfill
\begin{subfigure}[t]{0.32\linewidth}
    \centering
    \includegraphics[width=\linewidth]{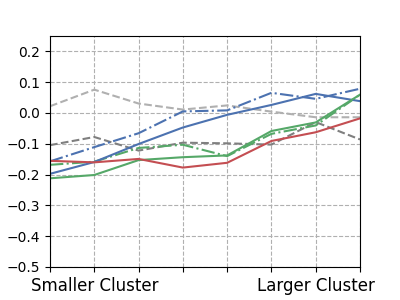}
    \caption{Qwen3-Emb.-4B}
\end{subfigure}
\hfill
\begin{subfigure}[t]{0.32\linewidth}
    \centering
    \includegraphics[width=\linewidth]{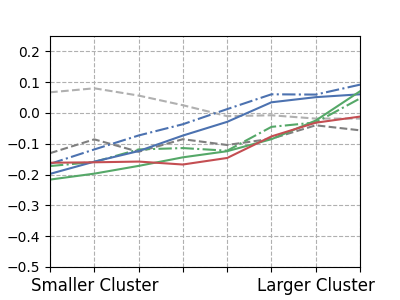}
    \caption{Qwen3-Emb.-8B}
\end{subfigure}

\caption{The Spearman's \(\rho\) correlation measured between the semantic entropy of each method and the pass@1 results when only the coder is involved.}
\label{fig:spearmanr_coder}
\end{figure}

Fig.~\ref{fig:spearmanr_coder} presents the Spearman’s \(\rho\) correlation between semantic entropy and pass@1 in the coder-only setting and all clustering methods show only weak correlation between 0.1 and -0.2. The light and dark grey dashed curves correspond to DBSCAN with a fixed  \(\epsilon\)and HDBSCAN with a fixed MCS, respectively. Both fixed clustering approaches exhibit weak correlation with the actual functional correctness of the generated code. The blue curves represent adaptive clustering based on the mean pairwise distance of the PDM, where the dashed and dotted variants correspond to approaches without additional scaling, and the solid variant incorporates the proposed scaling mechanism. Similarly, the green curves represent adaptive clustering based on the mean edge weight of the MST. Overall, semantic entropy derived from MST-based statistics demonstrates stronger correlation with functional correctness than approaches based solely on the PDM mean, and the inclusion of scaling further improves the correlation. The solid red curves correspond to semantic entropy computed using the MST mode as the adaptive reference value. This approach consistently achieves the strongest correlation across most clustering granularities, except in extremely fine-grained clustering settings where MST-mean-based approaches occasionally perform better.

\begin{figure}[htbp]
\centering

% Row 1
\begin{subfigure}[t]{0.32\linewidth}
    \centering
    \includegraphics[width=\linewidth]{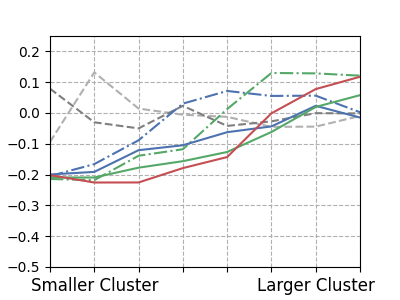}
    \caption{All-MiniLM}
\end{subfigure}
\hfill
\begin{subfigure}[t]{0.32\linewidth}
    \centering
    \includegraphics[width=\linewidth]{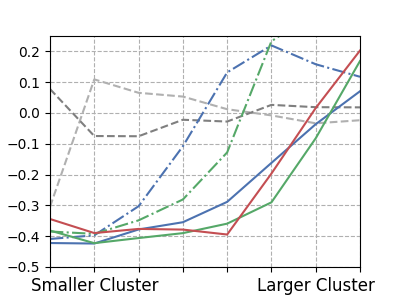}
    \caption{ModernBERT}
\end{subfigure}
\hfill
\begin{subfigure}[t]{0.32\linewidth}
    \centering
    \includegraphics[width=\linewidth]{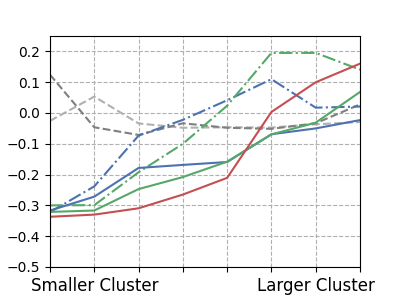}
    \caption{Nemotron}
\end{subfigure}

\vspace{0.5em}

% Row 2
\begin{subfigure}[t]{0.32\linewidth}
    \centering
    \includegraphics[width=\linewidth]{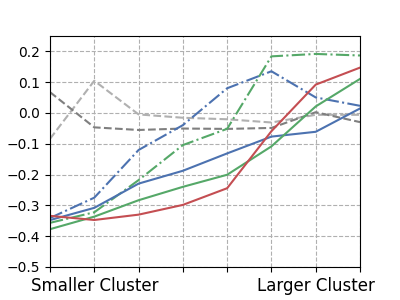}
    \caption{Qwen3-Emb.-0.6B}
\end{subfigure}
\hfill
\begin{subfigure}[t]{0.32\linewidth}
    \centering
    \includegraphics[width=\linewidth]{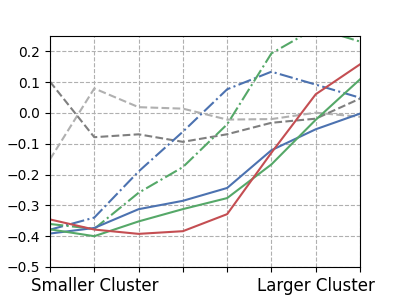}
    \caption{Qwen3-Emb.-4B}
\end{subfigure}
\hfill
\begin{subfigure}[t]{0.32\linewidth}
    \centering
    \includegraphics[width=\linewidth]{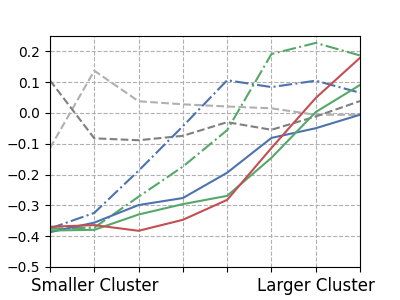}
    \caption{Qwen3-Emb.-8B}
\end{subfigure}

\caption{The Spearman's \(\rho\) correlation measured between the semantic entropy of each method and the pass@1 results when the analyst agent provides additional instruction to the coder}
\label{fig:spearmanr_analyst_coder}
\end{figure}

The adaptive clustering approaches exhibit stronger correlation when an analyst agent is introduced before the coder agent. As shown in Fig.~\ref{fig:spearmanr_analyst_coder}, nearly all adaptive clustering methods produce semantic entropy values with improved correlation to pass@1, whereas the fixed clustering approaches remain close to zero correlation. The relative ranking of the adaptive methods remains largely consistent, although the scaled PDM-mean approach now outperforms the MST-mean-based approach. Among all methods, the scaled MST-mode strategy achieves the strongest overall performance across nearly all settings, with the exception of Qwen3-Embedding-0.6B under extremely fine-grained clustering configurations, where alternative approaches perform better.

\begin{center} 
\begin{tcolorbox}[
  colback=gray!15,        % light gray background
  colframe=gray!50,       % gray frame
  arc=3mm,                % rounded corner radius
  width=0.9\textwidth,    % the box is 80% of text width
  left=4pt, right=4pt,    % horizontal padding
  top=4pt, bottom=4pt     % vertical padding
]

\textbf{Answer to RQ.2:} Task-specific adaptiveness is essential when clustering code samples represented as semantic embeddings. The strongest correlation between semantic entropy and functional correctness is consistently achieved by density-based clustering methods that use the scaled mode of MST edge weights as the neighborhood radius \(\epsilon\).
\end{tcolorbox}
\end{center}

\subsection{Comparison with LLM-based Semantic Entropy, Self-Evaluation Methods and Other Baselines}

To demonstrate the effectiveness of the proposed FASE measurement, this experiment compares it against the state-of-the-art semantic entropy approach based on bidirectional entailment using LLM judgments, as well as alternative entropy formulations derived from textual and structural equivalence. In addition, the evaluation includes widely used LLM self-evaluation baselines such as P(True) and majority voting, where functional correctness is determined according to predictions generated either by the coding LLM itself or by a committee of auxiliary LLMs. The evaluation considers two complementary perspectives: (1) Spearman’s \(\rho\)  coefficient, which measures the correlation between each uncertainty metric and the ground-truth pass@1 values, and (2) the Area Under the Curve (AUC) score, which evaluates the binary predictive capability of each metric in determining whether a task achieves pass@1 = 1.

\begin{table*}[htbp]
\centering
\caption{The Spearmnar's \(\rho\) correlation coefficient and AUC score for the baselines and FASE-related entropy using Qwen3-Embedding-8B in different workflow and coding LLMs. The entropy measurements are inverted by \(1-\text{entropy}\) to keep correlation positive, as lower entropy indicates greater consistency and better functional correctness. Best scores for each setting are highlighted.}
\label{tab:spearmanr_rocauc_all_metrics}
\setlength{\tabcolsep}{3pt}
\resizebox{\textwidth}{!}{%
\begin{tabular}{|l|cc|cc|cc|cc|cc|cc|cc|cc|}
\hline

% Row 1
\multirow{3}{*}{\textbf{Metric}} 
& \multicolumn{8}{c|}{\textbf{Coder Only}} 
& \multicolumn{8}{c|}{\textbf{Analyst + Coder}} \\ \cline{2-17}

% Row 2
& \multicolumn{2}{c}{Mistral}
& \multicolumn{2}{c}{CodeLlama}
& \multicolumn{2}{c}{DeepSeek}
& \multicolumn{2}{c|}{Qwen2.5}
& \multicolumn{2}{c}{Mistral}
& \multicolumn{2}{c}{CodeLlama}
& \multicolumn{2}{c}{DeepSeek}
& \multicolumn{2}{c|}{Qwen2.5} \\ \cline{2-17}

% Row 3
& \(\rho\) & AUC 
& \(\rho\) & AUC
& \(\rho\) & AUC
& \(\rho\) & AUC
& \(\rho\) & AUC
& \(\rho\) & AUC
& \(\rho\) & AUC
& \(\rho\) & AUC \\

\hline

Self P(True)  & 0.07 & 0.51 & 0.22 & 0.65 & 0.05 & 0.52 & 0.32 & 0.65 & 0.02 & 0.52 & 0.17 & 0.84 & 0.15 & 0.56 & 0.22 & 0.59 \\
Majority Voting  & 0.13 & 0.52 & 0.19 & 0.61 & 0.06 & 0.53 & 0.28 & 0.63 & 0.04 & 0.51 & 0.18 & 0.83 & 0.07 & 0.54 & 0.18 & 0.59 \\
Textual Entropy  & 0.06 & 0.64 & 0.18 & 0.74 & 0.24 & 0.66 & 0.36 & 0.73 & 0.26 & 0.6 & 0.24 & 0.78 & 0.4 & 0.76 & 0.41 & 0.67 \\
Structural Entropy  & 0.2 & \textbf{0.8} & 0.18 & 0.76 & 0.31 & 0.73 & 0.37 & 0.74 & 0.45 & 0.95 & 0.34 & 0.96 & 0.56 & 0.91 & 0.61 & 0.86 \\
Semantic Entropy  & 0.13 & 0.62 & \textbf{0.35} & \textbf{0.77} & 0.25 & 0.69 & 0.41 & 0.76 & 0.15 & 0.74 & \textbf{0.38} & 0.85 & 0.45 & 0.88 & \textbf{0.7} & \textbf{0.9} \\
\hline
Fixed MCS  & 0.04 & 0.63 & 0.12 & 0.69 & 0.19 & 0.63 & 0.02 & 0.52 & 0.07 & 0.57 & 0.02 & 0.51 & 0.22 & 0.61 & 0.13 & 0.53 \\
Fixed \(\epsilon\)  & 0 & 0.63 & 0 & 0.54 & 0.04 & 0.52 & 0.11 & 0.53 & 0.02 & 0.8 & 0.04 & 0.74 & 0.27 & 0.78 & 0.13 & 0.62 \\
FASE-PDM Mean  & 0.1 & 0.57 & 0.14 & 0.58 & 0.19 & 0.62 & 0.38 & 0.73 & 0.37 & 0.91 & 0.24 & 0.83 & 0.51 & 0.86 & 0.46 & 0.76 \\
FASE-MST Mean  & 0.31 & 0.73 & 0.22 & 0.68 & 0.39 & 0.73 & 0.48 & 0.75 & 0.37 & 0.91 & 0.26 & 0.83 & 0.51 & 0.86 & 0.46 & 0.76 \\
FASE-Scaled PDM Mean  & 0.31 & 0.73 & 0.24 & 0.74 & 0.39 & 0.73 & 0.48 & 0.75 & 0.37 & 0.93 & 0.28 & 0.84 & 0.51 & 0.86 & 0.46 & 0.76 \\
FASE-Scaled MST Mean  & 0.33 & 0.75 & 0.25 & 0.75 & 0.4 & 0.74 & 0.5 & 0.76 & 0.39 & 0.93 & 0.28 & 0.84 & 0.51 & 0.86 & 0.47 & 0.76 \\
FASE-Scaled MST Mode  & 0.34 & 0.75 & 0.25 & 0.75 & 0.4 & 0.74 & 0.5 & 0.76 & 0.39 & 0.93 & 0.3 & 0.89 & 0.51 & 0.86 & 0.47 & 0.76 \\
\hline
\textbf{\textit{FASE \(\cup\) Structural}}  & \textbf{0.34} & 0.77 & 0.28 & 0.75 & \textbf{0.4} & \textbf{0.74} & \textbf{0.5} & \textbf{0.76} & \textbf{0.47} & \textbf{0.96} & 0.34 & \textbf{0.98} & \textbf{0.59} & \textbf{0.91} & 0.61 & 0.87 \\

\hline
\end{tabular}
}
\end{table*}

The results, presented in Tab.~\ref{tab:spearmanr_rocauc_all_metrics}, demonstrate that FASE entropy consistently outperforms LLM-entailment-based semantic entropy in both correlation and predictive performance across code generation workflows involving either a standalone coder LLM or an analyst–coder pipeline. In particular, combining the semantic equivalence labels generated by FASE using scaled MST-mode clustering with structural equivalence classes through a \textit{union} operation produces the entropy measure with the strongest overall correlation and predictive capability, outperforming all baseline methods across the evaluated settings.

In the coder-only setting, when Mistral is used, all FASE-related variants achieve moderate correlation with functional correctness, although structural entropy attains the highest AUC score of 0.80. For CodeLlama, traditional semantic entropy outperforms the other metrics in both Spearman’s \(\rho\) and AUC, while FASE remains highly competitive. In the case of DeepSeek-Coder, FASE with scaled MST Mean/Mode achieves the strongest performance with a \(\rho\) of 0.40 and an AUC of 0.74. Similarly, for Qwen2.5-Coder, FASE with scaled MST Mean/Mode again yields the best overall results, reaching a \(\rho\) of 0.50 and an AUC of 0.76.

When an analyst agent is incorporated into the workflow, FASE \(\cup\) structural entropy provides the most accurate estimation of functional correctness for Mistral-generated outputs, achieving a \(\rho\) of 0.47 and an AUC of 0.96. For CodeLlama, the highest correlation is obtained using traditional semantic entropy with a \(\rho\) of 0.38, whereas FASE achieves the best predictive performance with an AUC of 0.98. FASE again produces the strongest results for DeepSeek-Coder, with a  \(\rho\) of 0.59 and an AUC of 0.91. For Qwen2.5-Coder, traditional semantic entropy achieves the best overall performance, reaching a \(\rho\) of 0.70 and an AUC of 0.90.

\begin{center} 
\begin{tcolorbox}[
  colback=gray!15,        % light gray background
  colframe=gray!50,       % gray frame
  arc=3mm,                % rounded corner radius
  width=0.9\textwidth,    % the box is 80% of text width
  left=4pt, right=4pt,    % horizontal padding
  top=4pt, bottom=4pt     % vertical padding
]

\textbf{Answer to RQ.3:} FASE and its structural hybrid variants outperforms existing state-of-the-art LLM-based semantic entropy methods by approximately 25\% in correlation and 8\% in AUC on average, achieving the strongest overall performance across most experimental settings involving all four selected LLMs and both the coder-only and analyst+coder workflows.
\end{tcolorbox}
\end{center}

\subsection{Computational Cost and Scalability Analysis}
One of the primary reasons that traditional semantic entropy is impractical in large-scale code generation evaluation is its computational cost from bidirectional entailment queries which scales in \(\mathcal{O}(|x|^2)\) as code sampling increases. Tab.~\ref{tab:time_cost_llm} records and compares the time required for all LLM-related tasks discussed in earlier evaluation steps, and Tab.~\ref{tab:time_cost_embedding} covers time costs associated with embedding models.

\begin{table}[htbp]
\centering
\caption{The time cost for LLM and embedding model tasks measured in seconds.}
\label{tab:time_cost}

\vspace{-0.3em}
% =========================
% Top Subtable
% =========================
\subcaption{The time cost of LLM-related tasks.}
\label{tab:time_cost_llm}
\vspace{-0.3em}

\begin{tabular}{|l|c c c c|}
    \hline
    \textbf{Observation Method} & \textbf{Mistral} & \textbf{Codellama} & \textbf{Deepseek-Coder} & \textbf{Qwen2.5-Coder} \\
    \hline
    Self-Evaluation & 3.34 & 25.4 & 3.74 & 2.72 \\
    Majority Voting & 31.86 & 9.8 & 31.46 & 32.48 \\
    Semantic Equivalence & 44.23 & 110.98 & 39.24 & 17.63 \\
    \hline
\end{tabular}

\vspace{1em}

% =========================
% Bottom Subtable
% =========================
\subcaption{The time cost of embedding-related tasks.}
\label{tab:time_cost_embedding}
\vspace{-0.3em}

\begin{tabular}{|l|c c|}
    \hline
    \textbf{Embedding Model} & \textbf{PDM Generation} & \textbf{MST Extraction} \\
    \hline
    All-MiniLM & 0.021 & 0.0000342 \\
    ModernBERT & 0.073 & 0.0000265 \\
    Nemotron & 0.327 & 0.0000265 \\
    \hline
    Qwen3-Emb.-0.6B & 0.112 & 0.0000265 \\
    Qwen3-Emb.-4B & 0.264 & 0.0000264 \\
    Qwen3-Emb.-8B & 0.357 & 0.0000264 \\
    \hline
\end{tabular}

\end{table}

The self-evaluation metric P(True) is relatively efficient to compute compared to more expensive semantic equivalence checks. For a batch of 10 self-evaluation predictions, Mistral, DeepSeek-Coder, and Qwen2.5-Coder require approximately 2–3 seconds, whereas CodeLlama requires substantially longer at 25.4 seconds. Majority voting, which relies on a committee of auxiliary LLMs excluding the coding model itself, incurs significantly higher computational cost, requiring approximately 30 seconds. In comparison, conducting bidirectional semantic equivalence checks across 10 code samples is considerably more expensive, taking 44.23 s, 110.98 s, 39.24 s, and 17.63 s for the respective models.

Tasks involving embedding models instead of LLMs are an order of magnitude faster than LLM-based operations. Computing the PDM for 10 code samples or functionality reviews requires only approximately 0.02–0.3 seconds. Among the evaluated embedding models, All-MiniLM-L6-v2 is generally the fastest, followed by GTE-ModernBERT, while the Nemotron and Qwen3-Embedding series are the slowest, consistent with their larger model sizes. The extraction of the MST from the PDM incurs negligible computational overhead and is effectively independent of the choice of embedding model.

\begin{center} 
\begin{tcolorbox}[
  colback=gray!15,        % light gray background
  colframe=gray!50,       % gray frame
  arc=3mm,                % rounded corner radius
  width=0.9\textwidth,    % the box is 80% of text width
  left=4pt, right=4pt,    % horizontal padding
  top=4pt, bottom=4pt     % vertical padding
]

\textbf{Answer to RQ.4:} The embedding models provide substantial efficiency gains, reducing the overhead to approximately 0.3\% of the traditional pairwise LLM-based equivalence checks on average.
\end{tcolorbox}
\end{center}

Overall, the experimental results provide strong evidence supporting the effectiveness and efficiency of the proposed framework. The results show that embedding-based representations meaningfully capture functional differences between code samples, and adaptiveness is crucial for reliable clustering, with density-based methods using scaled MST-PDM ratio. FASE consistently outperforms existing LLM-based semantic entropy methods and offers substantial efficiency benefits, reducing computational overhead to roughly 0.3\% of that required by traditional LLM-based equivalence checking. Collectively, these findings validate both the predictive strength and practical efficiency of FASE for scalable code quality estimation.

\section{Related Work}
\label{sec:related_work}
Several lines of research have explored the use of multiple LLM agents to enhance automated code generation. At the same time, considerable attention has been devoted to understanding their inherent limitations, particularly those arising from code quality issues, hallucination propagation, and strategies for mitigating cascading errors across collaborative agent workflows.

\noindent\textbf{Multi-agent Frameworks for Code Generation.} Recent advances in large language models (LLMs) have enabled autonomous AI agents for software development, motivating extensive research on agent architectures, collaboration, and evaluation~\cite{mohammadi2025evaluation,ferrag2025llm,liu2024large,he2025llm}. Prior studies highlight both the potential of these systems to automate complex software engineering tasks and the challenges they introduce, including error propagation, limited self-assessment, and coordination overhead~\cite{yehudai2025survey}. To improve robustness, multi-agent frameworks have been proposed to encourage specialization and collaboration among agents with distinct responsibilities~\cite{talebirad2023multi}. Several collaborative agent frameworks have shown promising results for code generation and repair. CodeCoR~\cite{pan2025codecor} employs specialized agents for prompt generation, coding, testing, and repair, while MetaGPT~\cite{hong2024metagpt} integrates human-inspired software engineering workflows into structured multi-agent prompting pipelines. Dong et al.~\cite{dong2024self} propose a self-collaboration framework with analyst, coder, and tester agents that substantially improves Pass@1 over single-agent generation. AdaCoder~\cite{zhu2025adacoder} further introduces adaptive planning and iterative debugging, achieving better generalizability, faster inference, and lower token consumption across diverse LLMs. Beyond code generation, multi-agent collaboration has also been applied to software requirement compliance verification through retrieval-augmented generation and human-in-the-loop feedback~\cite{das2025multi}. Collectively, these studies demonstrate the importance of adaptive coordination and reliability estimation in autonomous software engineering systems.

\noindent\textbf{Agent Hallucination and Uncertainty.} Hallucination in large language models has received significant attention, leading to extensive surveys and empirical studies on hallucination detection and mitigation~\cite{zhang2025siren,huang2025survey,zhang2025llm}. Prior work categorizes hallucinations across both natural language and code generation tasks, highlighting the challenge of identifying hallucinations without ground-truth answers. FEWL~\cite{wei2024measuring} introduces a principled hallucination metric that weights evaluator LLMs according to estimated expertise. In code generation, HalluCode~\cite{liu2024exploring} shows that current LLMs struggle to recognize and classify hallucinations effectively. CodeMirage~\cite{agarwal2024codemirage} further establishes one of the first comprehensive taxonomies of code hallucinations and introduces a dedicated benchmark for hallucination detection. Extending this direction, Liu et al.~\cite{liu2026beyond} propose a broader taxonomy covering multiple categories of hallucinations, their causes, and impacts across different models and benchmarks. Recent studies also explore improved detection methodologies. Yang et al.~\cite{yang2025advancing} propose a hybrid static–dynamic hallucination detection framework that substantially improves detection performance on multiple benchmarks. Foodeei et al.~\cite{foodeei2025semantic} investigate semantic uncertainty under different decoding strategies and show that structured reasoning can improve both semantic diversity and functional correctness. Collectively, these studies motivate the need for efficient semantic-level uncertainty estimation methods for reliable code generation systems.

\noindent\textbf{LLM Code Quality Assurance.} Several lines of research have explored software quality assurance for LLM-based systems. Self-confidence estimation methods such as P(True) and P(IK)~\cite{kadavath2022language} attempt to predict output reliability without ground-truth labels, while self-reflection frameworks improve consistency through iterative feedback and refinement~\cite{ji2023towards}. To reduce hallucinations in code generation, De-Hallucinator~\cite{eghbali2024hallucinator} iteratively retrieves project-specific API references to ground model predictions, significantly improving API usage and test generation quality. Repository-aware approaches such as \(A^3\)-CodGen~\cite{liao2024mathbf} further incorporate local, global, and third-party library context to reduce logical inconsistencies and improve code reuse. At the systems level, multi-agent software engineering frameworks increasingly adopt software engineering practices such as milestone-based collaboration assessment~\cite{zhu2025multiagentbench}, contribution-aware agent selection~\cite{liu2024dynamic}, and adaptive coordination strategies. Interactive workflows such as TICODER~\cite{fakhoury2024llm} improve code generation accuracy through iterative intent clarification and test-driven feedback. Recent work also explores LLM-assisted software testing and oracle generation. CANDOR~\cite{xu2026hallucination} proposes a multi-agent framework for automated unit test generation that combines specialized agents and consensus-based reasoning to mitigate hallucinations in generated test oracles, significantly improving oracle correctness and mutation scores over existing approaches. Empirical studies also examine the reliability and maintainability of LLM-generated code. Liu et al.~\cite{liu2024no} evaluate ChatGPT-generated programs across correctness, complexity, and security dimensions, highlighting issues related to non-determinism and vulnerability generation. Subsequent work~\cite{liu2024refining} shows that iterative self-repair and static-analysis-guided refinement can partially improve maintainability and correctness. In parallel, automated mutation testing~\cite{tip2025llmorpheus,wang2026mutation} and fuzz testing~\cite{cao2025program,chen2025traceawareness} demonstrate strong potential for validating LLM-generated code. Recent semantic representation learning approaches such as Tailor~\cite{liu2023learning} further emphasize the importance of semantic-level reasoning for reliable functional similarity detection and code quality assurance.

\noindent\textbf{Entropy of LLM output.} Several prior works have investigated uncertainty estimation in LLMs by accounting for code equivalence, recognizing that multiple surface forms may convey the same underlying meaning. To address this challenge, semantic entropy was proposed as an unsupervised uncertainty measure that aggregates model outputs at the level of meaning rather than token sequences~\cite{kuhn2023semantic}. These methods demonstrate robustness across tasks and datasets, generalize to unseen scenarios, and do not rely on prior task knowledge or labeled data~\cite{farquhar2024detecting}. In the context of code generation, structural entropy~\cite{song2025measuring} is specifically designed for source code represented as abstract syntax trees, providing an alternative approach for efficient code quality estimation. 

FASE provides a scalable uncertainty estimation mechanism that directly addresses a key limitation in current multi-agent code generation systems—namely, the lack of efficient and reliable measures for agent-level hallucination and error propagation. By bridging embedding-based semantic reasoning with entropy-driven evaluation, it complements existing LLM code quality assurance and entropy-based methods, offering a practical alternative for modeling uncertainty in complex multi-agent workflows.

\section{Threats to Validity}
\label{sec:threat_to_validity}

Despite its strong empirical performance, FASE inherits common limitations of embedding-based and clustering-driven approaches, as well as constraints of experiments:

\noindent{\textbf{Embedding Representation:}} A key limitation of textual and structural similarity measures is that the true semantics of a program may change drastically due to a single operator or variable modification. Although modern embedding models often employ attention mechanisms, embeddings of code snippets with opposite functionality may still exhibit high similarity, thereby undermining the validity of pairwise semantic comparisons. The combination of FASE and structural entropy through a union operation achieves the best overall performance in Tab.~\ref{tab:spearmanr_rocauc_all_metrics} because abstract syntax trees are highly sensitive to fine-grained token-level mutations that embedding-based semantic representations may overlook.

\noindent{\textbf{Prompt Engineering:}} The quality of prompts significantly influences LLM-generated outputs~\cite{khojah2025impact}. The experiments conducted in this study employ relatively simple prompting strategies. For code generation tasks, the models are provided with only basic coding instructions with a persona system prompt. For functional correctness and equivalence evaluation, a 2-shot prompting strategy is adopted, consisting of one positive example (functional/equivalent) and one negative example (failed/not equivalent). The extent to which FASE generalizes under more sophisticated prompting strategies remains an open question.

\noindent{\textbf{Code Generation Scale:}} Due to computational resource constraints, this study evaluates only open-source models with 7B parameters. The inclusion of additional models with broader architectural diversity and varying parameter scales may influence the observed performance and generalizability of FASE.

\section{Conclusion \& Future Directions}
\label{sec:conclusion}

This research explores the application of FASE entropy in code generation and demonstrates its effectiveness in estimating the functional correctness of generated code in the absence of ground-truth test cases. Given the impractical computational cost of LLM-based bidirectional equivalence checks, this research offers a scalable alternative that combines code semantic and structural consistency. Extensive evaluations across four widely used code LLMs and four embedding models show that this approach achieves high correlation and predictive performance while significantly reducing computation time. By enabling efficient, ground-truth–free monitoring of code quality, the proposed FASE entropy supports multi-agent code generation workflows with reduced risk of error propagation and hallucination.

Several directions remain for future work. First, the evaluation can be expanded to include a broader range of foundation LLMs, embedding models, and code generation benchmarks to further validate the robustness and generalizability of FASE. Second, FASE does not explicitly distinguish cases where all generated solutions consistently drift away from the intended functionality with low entropy. Incorporating an additional dimension of semantic alignment with task requirements may improve the detection of functionally incorrect yet semantically consistent outputs. Finally, FASE can be integrated directly into adaptive multi-agent decision-making frameworks, where uncertainty estimates dynamically guide agent selection, workflow orchestration, refinement strategies, or verification depth.

% check ACM reference style
\bibliographystyle{acm}
\bibliography{sample-base}
\end{document}